\documentclass[10pt,twocolumn,letterpaper]{article}
\usepackage{amscd}
\usepackage{amssymb}
\usepackage{times}

\font\elvbf  = ptmb scaled 1100

\def\R{{\mathbb R}}

\def\e{{\epsilon}}

\newcommand{\abs}[1]{\left\vert#1\right\vert}
\def\norm#1{\left\Vert#1\right\Vert}

\newtheorem{thm}{Theorem}[section]

\newtheorem{prop}[thm]{Proposition}
\newtheorem{defin}[thm]{Definition}

\def\sect #1{{\stepcounter{section} \vskip .35cm
\noindent{\large\bf\arabic{section}. #1}
 \vskip .35cm
}}
\def\section*#1{{\sect #1}}

\def\subsect #1{{\stepcounter{subsection} 
\noindent{\elvbf\arabic{section}.\arabic{subsection}. #1}
 \vskip .3cm
}}

\begin{document}
\title{\Large\bf A Geometric Framework for Modelling Similarity 
Search\footnote{Accepted, to appear in Proc. Intern. Workshop on
Similarity Search (IWOSS'99), Florence, Sept. 1999.}
\\[.2cm]
{\large\rm Vladimir Pestov}}
\author{\large
\it
School of Mathematical and Computing Sciences, \\
\large\it Victoria University of
Wellington, \\
\large\it P.O. Box 600,
Wellington, New Zealand
\footnote{Permanent address.} 
\\[.1cm]
\and
\large\it Computer Sciences Laboratory, RSISE,\\
\large\it Australian National University, \\
\large\it Canberra, ACT 0200, Australia
\footnote{Visiting ACSys Fellow, July 1 -- Dec. 31, 1999.} 
\and
\large\it vova@mcs.vuw.ac.nz \\
\large\it http://www.vuw.ac.nz/$^\sim$vova}
\date{}
\maketitle
\thispagestyle{empty}

\vskip .35cm

\centerline{\large\bf Abstract}
   
{\it
\noindent 
We suggest a geometric framework for
modelling similarity search in large and multidimensional
data spaces of general nature, formed
by the concept of the similarity workload, which is 
a probability metric space $\Omega$ (query domain) with a
distinguished finite subspace $X$ (dataset), together with an assembly
of concepts, techniques, and results from metric geometry. 
As some of the latter are being currently reinvented
by the database community, it seems desirable to try and
bridge the gap between database research and the relevant work already
done in geometry and analysis.}

\vskip .3cm

\sect{Introduction}

Mathematical modelling of similarity search is still
very much in its infancy, and the 
modest aim of this paper is to spot a few 
mathematical structures clearly emerging 
in the present practice of similarity search
and point in the direction of some relevant and
well-established concepts, ideas, and techniques which belong to 
geometric and functional analysis and appear to be relatively
little known in the database community.

For the most part, there is a long way to go
before (and if) the outlined 
ideas and methods are put into a workable shape and
made relevant to the concrete needs of similarity search. This is
certainly the case with the phenomenon of concentration of measure
on high-dimensional structures, which might potentially 
have the greatest impact of all
on both theory and practice of similarity
search, through offering a possible insight into the nature of the
curse of dimensionality. However, in some other instances
the established mathematical methods can be used directly
and, we believe, most profitably, in order to improve the
existing algorithms for similarity retrieval based on {\it ad hoc,}
though often highly ingenious, mathematical techniques.
This could well be the case with the technique of metric
transform as applied to histogram indexing for image search
by colour content. Even here the gap separating
theory and practice of database research (discussed in \cite{Papa}
in a highly colourful manner) has to be bridged yet. 
Nevertheless, as the size of datasets grows
exponentially with time, attempts to understand 
the underlying, very complex, geometry of similarity search through
joint efforts of mathematicians and computer scientists seem to have
no credible alternative. 

\sect{Data sets and metrics}
\subsect{Data structures}
An undisputable --- though 
often downplayed in theoretical analysis ---
fact is that a query point, $x^\ast$, need not
belong to an actual dataset, $X$. 
This is why we make a distinction between
the collection of query points (domain) and the actual dataset. 
A {\it domain} is a 
metric space, $\Omega=(\Omega,\rho)$, whose elements are {\it query points,}
and the metric $\rho$ is the {\it dissimilarity measure.} 
An actual {\it dataset} (or {\it instance}),
$X$, is a finite metric subspace of $\Omega$. 

We have borrowed the concept of a domain from \cite{HKP}, where it is
defined as `a set such as $\R^d$ together with methods such as
$x$-component, order, etc.' 
Even if dissimilarity measures not satisfying the
triangle inequality are sometimes considered 
\cite{FS}, namely metrics, or
else pseudometrics (for which
the condition $(\rho(x,y)=0)\Rightarrow (x=y)$ is dropped),
are of overwhelming importance.
Many metric spaces appearing in this context
are non-Hilbert, e.g. the Hamming cube $\{0,1\}^n$ equipped with the
string edit distance, or $l_1$. 
In fact, it is hardly possible to come up with any 
{\it apriori} restrictions
that distinguish metric spaces `relevant for applications' from those
that are not.

\vskip .35cm

\subsect{Similarity queries and indexing}
Two major types of similarity queries are range queries and nearest
neighbour queries.
Let $x^\ast\in\Omega$ be the query centre. 
A generic $\e$-{\it range query}
is of the form: given an $\e>0$,
find all $x\in X$ with $\rho(x,x^\ast)<\e$.
A generic {\it $k$-nearest neighbour  {\rm (}$k$-NN{\rm)}
query} is of the form: given a
natural $k$ and an $x^\ast\in\Omega$, find $k$ elements 
of $X\setminus\{x^\ast\}$ closest to $x^\ast$ in the sense of 
metric $\rho$.

A $k$-NN query can be reduced 
to a series of range queries with varying radii $\e>0$ chosen by
some sort of a binary procedure. 

A general {\it index structure} \cite{HKP}
on a dataset $X$ is just any
cover $\Gamma$ of $X$ (that is, $\cup\Gamma=X$)
with a collection of {\it blocks} $A\in\Gamma$
of uniform and `manageable' size.

\begin{defin}
A {\it hierarchical tree index structure} on a set $X$ is a family 
$\Gamma=\{A_t\}_{t\in T}$ of subsets of $X$ (blocks) indexed
with elements (nodes) of a finite tree $T=(T,\leq)$ in such a way that the 
following are satisfied.
\begin{enumerate} 
\item For the root $0\in T$, $A_{0}=X$.
\item If $t\in T$ and $t_i$ are descendants of $t$, then
the sets $A_{t_i}$ cover $A_t$.
\end{enumerate} 
\label{tree}
\end{defin}

This scheme apparently includes as particular cases 
$k$-d tree, metric tree, {\it vp-}tree,
{\it gh-}tree, GNAT, M-tree, pyramid technique, etc. (See e.g.
\cite{pyr, Brin, CPZ, U1, WSB} and references therein.)
\vskip .35cm

\subsect{\label{processing}Processing range queries}
To process a range query of radius $\e>0$ centred at
$x^\ast\in\Omega$ using a hierarchical tree indexing structure $\Gamma$
on $X$, one employs the following algorithm. Below
\[{\mathcal O}_\e(A)=\{x\in \Omega\colon \rho(x,a)<\e ~~
\mbox{for some $a\in A$}\}\]
is the $\e$-neighbourhood of $A$ in $\Omega$.

\begin{enumerate}
\item Set $t=0$.
\item Set $A=A_t$.
\item If $A$ is a leaf, use exhaustive search to find all $x\in A$
with $\rho(x,x^\ast)<\e$.
\item Else, if it can be certified that $x^\ast\notin{\cal O}_\e(A)$,
prune the sub-tree descending from the internal node $t$.
\item Else, for every $i=1,2,\dots,k$, where $t_1,t_2,\dots,t_k$
are descendants of $t$, do: set $t=t_i$ and go to 2.
\end{enumerate}

If we had means to certify at each step 
that $x^\ast\in{\cal O}_\e(A)$, then the algorithm could be modified so
as to return one of the $\e$-neighbours 
in time $O(h)$, where $h$ is the height of the tree $T$
(typically, $O(\log n)$, with $n$ the number of objects in the dataset),
by traversing down one branch and selecting at each step
an arbitrary node $t$ such that $A$ and ${\cal O}_\e(x^\ast)$ have a
non-empty intersection.

Unfortunately, even if the possibility of such certification 
was (implicitely) assumed by some authors, it is 
computationally unfeasible.
Instead, the following technique is employed. 

A function $f\colon \Omega\to\R$ is called
{\it $1$-Lipschitz} if \[\abs{f(x)-f(y)}\leq\rho(x,y)\] for each
$x,y\in \Omega$. For each $t\in T$, choose
computationally inexpensive 1-Lipschitz functions $f_t$
and numbers $a_t,b_t>0$ such that
$f_t(A_t)\subseteq [a_t,b_t]$. If $x\in {\cal O}(A_t)$, then the
Lipschitz-1 property of $f_t$ implies $f_t(x)\in
(a_t-\e,b_t+\e)$. Thus, the property
$f_t(x)\notin (a_t-\e,b_t+\e)$ is a certificate for 
$x\notin A_t$. Yet, the condition $f_t(x)\in (a_t-\e,b_t+\e)$
does not allow one to make a conclusion about whether or not
$x$ is in ${\cal O}(A_t)$, and {\it every} such node has to be 
followed through.

An example of such kind is
the {\it distance} function $d_t$ from some $v_t\in\Omega$, called
a {\it vantage point} (for the node $t$):
\[d_t\colon x\mapsto d(v_t,x),\]
where $a=\min\{d_v(x)\colon x\in A\}$ and
$b=\max\{d_v(x)\colon x\in A\}$ are the corresponding 
precomputed constants. 

In fact, every subset $A\subseteq X$ 
admits an {\it exact} Lipschitz-1 certification function ---
namely, the distance from $A$:
\[d_A\colon x\mapsto d(x,A)=\inf\{\rho(x,a)\colon a\in A\},\]
with constants $a=b=0$.
However, the function $d_A$ is normally far too expensive
computationally to be used.

\sect{Changing the distance}
\subsect{\label{replace}The first complexity issue}
Processing a similarity query requires a large number of
computations of the values of certification functions, typically
distances between points. Often performing even a single 
computation of the value of the original dissimilarity measure,
$\rho$, is time-consuming.
This is why the following technique is often applied --- so often,
in fact, that we consider it to form a major
component of the abstract geometric framework. 
\smallskip

1. The `exact' dissimilarity measure $\rho$ on $\Omega$ is replaced
with a computationally cheaper distance $d$.

2. For given $x^\ast\in\Omega$ and 
$\e>0$ one chooses a $\delta>0$ such
that for every $x\in X$, the condition 
$\rho(x,x^\ast)<\e$ implies $d(x,x^\ast)<\delta$. Now,
instead of processing the $\e$-range query in
$(\Omega,\rho,X)$ centred at $x^\ast$, one
processes the $\delta$-range query in
$(\Omega, d,X)$ centred at $x^\ast$.

3. For each returned $x\in X$, the condition $\rho(x,x^\ast)<\e$
is verified and the false hits discarded.
\smallskip

Dimensionality reduction and the projection 
search paradigm
(see e.g. \cite{NN}) are examples of the above technique.
If $\Omega$ is a metric subspace of a high-dimensional Euclidean space
$l_2^N$
(that is, $\rho(x,y)=\norm{x-y}_2$, where 
$\norm\cdot_2$ is the Euclidean distance)
and $\pi$ denotes the projection onto a Euclidean subspace
$l_2^n\subset l_2^N$ of a 
lower dimension $n<N$, then $d$ is defined by the formula
\[d(x,y)=\norm{\pi(x)-\pi(y)}_2.\]
More generally, this is the case with every distance $d$ used
for {\it prefiltering} \cite{SW}.
In this case, $d(x,y)\leq\rho(x,y)$ for all $x,y$. 
\vskip .35cm

\subsect{Metric transform}
A rich source of new metrics $d$ on $X$
leading to the same nearest neighbour graph \cite{EPY} as the original
metric $\rho$ is the classical construction of metric transform
\cite{DL}.
Let $(X,\rho)$ be a metric space and let $F\colon\R_+\to\R_+$
be a concave non-decreasing function satisfying $F(0)=0$.
A {\it metric transform} of $X$ 
by means of $F$ is a pair $F(X)=(X,F(\rho))$, where $F(\rho)$ is
a metric on $X$ defined by
$F(\rho)(x,y)=F(\rho(x,y))$. 

The theory of metric transform is fairly advanced. 
Often
the metric transform of a non-Euclidean metric space turns out
to be Euclidean and therefore computationally simple. At the same
time, the metric transform itself can be performed at the
database population stage.

\vskip .35cm

\subsect{Example: quadratic distance}
If $C=\{c_1,\dots,c_n\}$ is a finite set,
then a {\it histogram} on $C$ is an element of the convex hull of $C$,
which we will denote by
$P(C)$, that is, a linear combination
$\sum_{i=1}^n\lambda_ic_i$, $\lambda_i\geq 0$,
$\sum_{i=1}^n\lambda_i=1$. An example we will have in mind is that
of a {\it colour histogram,} showing the colour content of an image.
Here $C$ is a {\it colour space}, which is typically 
a convex subset of a low-dimensional
Euclidean space equipped with the induced distance, $\rho_C$, and
in practice replaced with
a finite metric subspace through a suitable colour segmentation
procedure. Histograms over $C$ are exactly the
distributions of {\it image functions} taking
values in $C$. The most natural distances on the space
of histograms, $P(C)$, are {\it probability metrics}
\cite{Rach},
in particular the well-known {\it Kantorovich distance,} 
defined for each $\mu_1,\mu_2\in P(C)$
by:
\begin{eqnarray}
\hat \rho(\mu_1,\mu_2)&= & \inf \left\{ \sum_{i,j=1}^n 
 \vert \lambda_{ij}\vert
\rho(c_i,c_j)    \colon  ~\lambda_{ij}\geq 0,\right. \nonumber \\
& & 
\left.
\mu_1-\mu_2=\sum_{i,j=1}^n 
\lambda_{ij}\left(c_i-c_j\right)\right\}.
\label{free}
\end{eqnarray}
The Kantorovich metric has the following `universal property.'
Recall that a map is {\it affine} if images of segments of straight
lines are again such.

\begin{prop} 
Every non-expansive mapping $f$ from a finite metric
space $C$ to a normed space $E$ extends to a unique affine mapping
$\tilde f\colon P(C)\to E$, which is 1-Lipschitz with respect to
the Kantorovich distance.
\label{mapping}
\end{prop}

The Kantorovich distance is 
computationally expensive. 
The QBIC project \cite{HSEFN} employs instead 
the so-called quadratic distance,
which is Euclidean and obtained by means of metric transform 
(though neither fact was realized by its inventors
and a full advantage of them never taken).

A {\it quadratic distance} \cite{HSEFN}, $d$, on the convex 
hull $P(C)$ of a finite set $C=\{c_1,c_2,\dots,c_n\}$ is the
distance determined by the inner product 
\begin{equation}
(x,y)=xAy^t
\label{inn}
\end{equation}
on the linear space spanned by $C$, where
$A$ is a symmetric $n\times n$-matrix satisfying a certain positive
(semi)definiteness condition.
Every mapping $f$ from $C$ to a Hilbert space $\mathcal H$ extends
to an affine map $\bar f\colon P(C)\to {\mathcal H}$, and the distance
$d(x,y)=\norm{\bar f(x)-\bar f(y)}$ is easily verified to be quadratic.
Moreover, one can prove that every quadratic distance on $P(C)$
is obtained in this way.
If now $C$ is equipped with a metric and the mapping $f\colon C\to 
{\mathcal H}$ is an isometric embedding 
of some metric transform of $C$, then
one obtains quadratic distances of the type used in the QBIC
project \cite{HSEFN}.  
One of the two main distances of this kind used in 
\cite{HSEFN} was determined 
by the matrix $a_{ij}=1-d_{ij}$, where
$d_{ij}$ are normalized Euclidean 
distances between elements of the colour space $C$. 
Using \cite{BG}, one can prove that this distance 
is obtained (up to a scalar multiple) in the above way
via applying to the Euclidean colour space the metric 
transform $F(t)=\sqrt t$. In addition, $C$ with this distance is
contained in the unit sphere of a Euclidean space.
\vskip .35cm

\sect{Geometry vs complexity}

\subsect{Measure concentration phenomenon}
From now on we will equip the query domain $\Omega$ 
with a probability Borel 
measure, $\mu$. 
(That is, $\mu$ is a sigma-additive measure with
$\mu(\Omega)=1$, defined on all sets that can be obtained from
open balls through countable unions, intersections, and complements.)
We will think of $\mu$ as reflecting the query distribution.

The quadruple $(\Omega, d,\mu, X)$ will be called a
{\it similarity workload.} 

Recall that a pair formed by a metric space
$(\Omega,\rho)$ and a probability measure $\mu$ on it is called
a {\it probability metric space.}
The {\it concentration function,}
$\alpha=\alpha_\Omega$, of a probability metric space $\Omega$ is defined 
by
\begin{eqnarray*} 
\alpha_\Omega(\e)=1-\inf\left\{\mu\left({\mathcal O}_\e(A)\right)
\colon A \mbox{ is Borel, }
\mu(A)\geq \frac 1 2\right\},
\end{eqnarray*}
for each $\e>0$ and $\alpha_\Omega(0)=1/2$. 
It is a decreasing function in $\e$. If $\alpha$ decreases sharply,
then most points of $\Omega$ are close to every subset
$A\subseteq\Omega$ containing at least a half of all points.
Most `naturally occuring' high-dimensional
probability metric spaces have
sharply decreasing concentration functions.
High-dimensional spheres, balls,
Hamming cubes, Euclidean cubes, Euclidean spaces equipped
with the Gaussian measure, groups of unitary matrices and numerous
other objects all have concentration functions 
not exceeding $C_1\exp(-C_2n\e^2)$, where $n$ is
the dimension and $C_1,C_2>0$. This observation is known
as the {\it phenomenon of
concentration of measure on high-dimensional structures}
\cite{GrM,M,Ta1}. Very large random structures, 
such as spin glasses, also exhibit this sort of behaviour
\cite{Ta2}. 

Assuming a typical multidimensional dataset to have a sharply
decreasing concentration function allows
one to explain at least some aspects of the dimensionality curse.
At the same time, such an assumption on the geometry of data
is much broader than that of uniformity and independence type
(cf. \cite{WSB}).
For an approach based on the concept of query instability,
proposed in
\cite{BGRS}, we refer the reader to our note \cite{P}. 
(This is why we do not discuss the paper \cite{BGRS} here.)

Notice that in some concrete large datasets the distribution
density of the distance functions (which are 1-Lipschitz) is known
to sharply peak near one value. And this is exactly the kind of
behaviour one would expect in the presence of concentration property,
in view of the following well-known result.

\begin{prop}
\label{conc}
Let $f\colon \Omega\to \R$ be a 1-Lipschitz function, and
denote by $M$ a median of $f$, that is,
a real number with 
$\mu(\{x\in X\colon f(x)\leq M\})=\mu(\{x\in X\colon f(x)\geq M\})$.
Then for every $\e>0$,
$\mu\left(f^{-1}(M-\e,M+\e) \right)\geq 1-2\alpha(\e)$.
\end{prop}

What is still missing, is a series of
computational experiments allowing
one to estimate the concentration functions of
large real datasets. 
\vskip .35cm

\subsect{False hits and metric entropy}
Within the outlined paradigm, the phenomenon of concentration of
measure can contribute towards the curse of dimensionality
through a massive amount of false hits
returned by similarity search algorithms.

To illustrate this on a simple example,
consider the projection search paradigm where
$\Omega\subset\R^N$ and $\pi\colon \R^N\to \R$ is 
the projection on the chosen coordinate axis (that is, $n=1$).
It follows from Proposition \ref{conc}
that for some $x^\ast\in\Omega$ 
query points $x$ with the property $\abs{\pi(x)-\pi(x^\ast)}<\e$
form a set of measure $\geq 1-2\alpha(\e)$.
If the concentration function of $\Omega$
is exponential in dimension, it means that the NN search along a 
single coordinate will return all datapoints located in a region of
$\Omega$ having nearly full measure.

To formulate a general result,
assume that the query domain $\Omega$ is compact. 
(This assumption does not seem to be at all restrictive.)
For an $\e>0$ denote by $N_\Omega(\e)$ the minimal number of open
$\e$-balls needed to cover $\Omega$. (Usually instead of this
quantity one considers its base $2$ logarithm, denoted by 
${\mathcal H}_\e(\Omega)$ and called the {\it $\e$-entropy}  
of $\Omega$.)
Let $\alpha$ denote the concentration function  
of the probability metric space $(\Omega,\rho,\mu)$.

\begin{prop} 
\label{two}
Let $\rho$ and $d$ be two distances on the same
probability space $(\Omega,\mu)$. 
Then there is a query point $x^\ast\in\Omega$ with 
the following property.
Let $\e,\delta>0$ be such 
that $(\rho(x,y)<\e/3)\Rightarrow (d(x,y)<\delta /3)$
for all $x,y$. Then the open ball formed
in $(\Omega,d)$ of radius $\delta$
has measure \[\mu({\mathcal O}_\delta(x^\ast))\geq 1- 
\alpha\left(\frac\e 3-\alpha^{-1}(N_{(\Omega,d)}(\e/3))
 \right).\]
\end{prop}

If now the query domain $(\Omega,\rho)$ 
has the sharply decreasing concentration function, while
the `capacity' of the
adjusted metric space, $(\Omega,d)$, is low, then 
for some query
point, $x^\ast$, every $d$-ball centred at $x^\ast$ and
containing the $\rho$-ball of radius $\e$,
would contain most of the query points and therefore quite 
probably a large amount of data points as well.
The transition from $\rho$ to $d$ results in an highly undesirable
`blow-up' of the mass of the query domain. 

The {\it trade-off} between the complexity of computing the 
distance $d$ and the suitably interpreted 
`capacity' of the space $(\Omega,d)$ could
be an important issue in optimising algorithms
for similarity search. Given a similarity workload $(\Omega,\rho,\mu,X)$,
does there exist an approximate distance $d$ which is computationally
simple and yet leaves ample space for the dataset to fit in $(\Omega,d)$
without `overcrowding'?
\vskip .35cm

\subsect{Example: average colour prefiltering}
In our simplified model the colour space $\mathcal C$ will
form an equilateral triangle with side one
having R, G, B as its vertices and equipped with the Euclidean distance.
It is segmented and replaced with a finite subset $C$ arranged in
a hexagonal lattice.
By $k$ we will denote the number of pixels in the image frame.
A {\it colour image} is an arbitrary function
({\it picture function}) from 
$\{1,2,\dots,k\}$ to $C$. 
The set $C^k$ of all colour images is given the normalized counting
measure $\mu_\sharp$.
For every image $x\in C^k$ denote by $\sigma(x)\in P(C)$ 
its colour histogram. One can prove that $\sigma$ is a 1-Lipschitz map.
Denote $P_k(C)=\sigma(C^k)$ and endow $P_k(C)$
with the direct image measure $\sigma_\ast(\mu_\sharp)$,
that is, the measure
of a subset $A\subseteq P_k(C)$ equals $\abs{\sigma^{-1}(A)}/n^k$.
The probability measure space $(P_k(C),\hat\rho,\mu_\sharp)$ forms the
query domain for {\it image query by colour content.}
It follows from results of \cite{Sch} that the 
concentration function, $\alpha$, of 
$(P_k(C),\hat\rho,\sigma_\ast(\mu))$ satisfies
\[\alpha(\e)\leq \frac 12\exp\left(-\frac{\e^2 k}4\right).\]
The embedding $C\hookrightarrow {\mathcal C}$ extends, by 
Proposition \ref{mapping}, to an affine
1-Lipschitz map $i\colon P(C)\to {\mathcal C}$, which is 
called the {\it average colour
map} in \cite{HSEFN} and
used for prefiltering in the QBIC project.

Using the same technique as in Proposition \ref{two}, one can show
that if $x^\ast$ is any colour histogram whose average colour is 
$\frac 1 3(R+G+B)$, then 
the $\e$-range query by colour content, preprocessed
using the average colour distance, will return all images contained
in a region of $C^k$ having measure at least 
\[1- 2\exp\left(-\frac{\e^2 k}{8}\right).\]
For example, if $k=100\times 100$ and $\e=0.1$,
the measure of the above region exceeds $0.99999$. 
Notice at the same time that 
the area of the corresponding open ball inside the colour space
$\mathcal C$ is at most $0.073$
of the area of the triangle.

\sect{Conclusion}

To quote \cite{HKP}, 
``What seems to be needed is a kind of {\it theory of indexability,} a
mathematical methodology which, in analogy with tractability,
would evaluate rigorously the power and limitations of the indexing
techniques in diverse contexts.''

This note is a fragment of what might develop into
{\it geometric theory of indexability} --- and, most importantly, 
an invitation for collaboration as well.

\end{document}